\documentclass[a4paper]{jpconf}
\usepackage{graphicx}

\def\timesbox{\hbox{$\scriptscriptstyle\times$}}
\def\ant{ {{\lower 1ex  \timesbox} \atop {\raise 1.5ex  \timesbox}}}
\newcommand{\Zop}{{\hbox{ Z\kern-1.6mm Z}}}
\newcommand{\beq}{\begin{equation}}
\newcommand{\eeq}{\end{equation}}
\newcommand{\bea}{\begin{eqnarray}}
\newcommand{\eea}{\end{eqnarray}}
\newcommand{\ra}{\rangle}
\newcommand{\la}{\langle}
\newcommand{\lt}{\left}
\newcommand{\rt}{\right}
\newcommand{\Iop}{\relax{\rm I\kern-.18em I}}
\newcommand{\one}{{\hbox{ 1\kern-1.2mm l}}}
\newcommand{\T}{{\cal T}}

\newcommand{\dt}{\delta}
\newcommand{\del}{\partial}

\newcommand{\eps}{\epsilon}

\newcommand{\s}{\sigma}

\newcommand{\xh}{\hat x}
\newcommand{\ph}{\hat p}
\newcommand{\pih}{\hat \pi}

\begin{document}

\hfill\vbox{\hbox{IMSc/2010/3/4}}


\title{String worldsheet theory in hamiltonian framework and background independence}

\author{Partha Mukhopadhyay}

\address{The Institute of Mathematical Sciences, C.I.T. Campus,
  Taramani, Chennai 600113, India} 

\ead{parthamu@imsc.res.in}

\begin{abstract}
We analyze exact conformal invariance of string worldsheet theory in
non-trivial backgrounds using hamiltonian framework. In the first part
of this talk we consider the example of type IIB superstrings in 
Ramond-Ramond pp-wave background. In particular, we discuss the
quantum definition of energy-momentum (EM) tensor and two methods of 
computing Virasoro algebra. One of the methods uses dynamical 
supersymmetries and indirectly establishes (partially) conformal 
invariance when the background is on-shell. We discuss the problem of 
operator ordering involved in the other method which attempts to
compute the algebra directly. This method is supposed to work for
off-shell backgrounds and therefore is more useful. In order to
understand this method better we attempt a background independent
formulation of the problem which is discussed in the second half of the
talk. For a bosonic string moving in an arbitrary metric-background
such a framework is obtained by following DeWitt's work 
(Phys.Rev.85:653-661,1952) in the context of particle quantum mechanics.
In particular, we construct certain background independent analogue of quantum
Virasoro generators and show that in spin-zero representation they
satisfy the Witt algebra with additional anomalous terms that vanish
for Ricci-flat backgrounds. We also report on a new result which
states that the same algebra holds true in arbitrary tensor
representations as well.  
 
\end{abstract}

\section{Introduction and summary}
\label{s:intro}

This talk will be based on the papers \cite{pm08, pm0902, pm0907,
  pm0912} and certain new results in \cite{pm10}. In \cite{pm08,
  pm0902, pm0907} we try to understand exact conformal invariance of
string worldsheet theory in certain pp-wave backgrounds using
hamiltonian framework\footnote{Two-dimensional non-linear sigma model
  is usually studied using lagrangian framework
  \cite{lagrangian}. Hamiltonian framework has also been used to
  some extent earlier in \cite{hamiltonian, diakonou90} (see also
  \cite{jain87}).}. Usually backgrounds that 
are nearly flat are considered for such
analysis. The reason is that for flat background the theory is
exactly solvable and the relevant vacuum is known. One can then define
the quantum EM tensor in the weak background field approximation by
normal ordering the interaction terms with respect to the
flat-space-vacuum. The pp-wave background on the other hand is not
close to flat and therefore the corresponding worldsheet theory is
strongly interacting. We will discuss this problem in section 
\ref{s:pp-wave}. 

The end result of the analysis in section \ref{s:pp-wave} will show
that there exists an operator ordering ambiguity in the computation of
the Virasoro anomaly if one tries to calculate them directly using the
quantum EM tensor. In order to have a better understanding of the
situation we seek a framework where the computations can be done in a
background independent way. Such a framework was developed by DeWitt
in \cite{dewitt52} for particle quantum mechanics. As will be
discussed in section \ref{s:back-indep}, the classical worldsheet
theory can be interpreted to be a theory of a particle moving in an 
infinite-dimensional curved background subject to certain potential.
It was shown in \cite{pm0912} that the background independent analogue
of the quantum Virasoro generators constructed in the sense of DeWitt
satisfy, in spin-zero representation, the Witt algebra with additional
anomalous terms that vanish for Ricci-flat backgrounds. We also
discuss how this might possibly suggest a resolution of the ordering
problem for the pp-wave.

In order to understand the implications of the Ricci-flatness
condition mentioned above various other questions regarding this
construction need to be answered. One such question, namely extending 
the analysis to higher rank tensor representations, has been considered in
\cite{pm10}. The results show that the same Ricci-flatness condition
holds true in general. We will discuss this in section \ref{s:back-indep}.
We end with some final remarks in section \ref{s:final}.

\section{Hamiltonian framework for worldsheet theory in pp-wave}
\label{s:pp-wave}

Here we will consider a pp-wave background, with or without R-R flux,
such that the only non-trivial component of the metric is\footnote{In
  absence of R-R flux, the pp-wave should correspond to an exact
  CFT whenever it is Ricci-flat (i.e. vanishing transverse laplacian
  of $K$) \cite{amati88}. We will consider a
  particular case of constant R-R flux in type IIB string theory, 
in which case conformal invariance should require the Ricci tensor to be
  proportional to the square of the flux strength \cite{blau01}.},
\bea
G_{++}= K(\vec x)~,
\eea
where the vector sign refers to the transverse directions $x^I$,
$I=1\cdots D-2$, $D$ being $26$ or $10$ depending on bosonic/superstring theory.
A plane-wave is a pp-wave for which $K(\vec x)$ is quadratic in $x^I$.

Proving conformal invariance for the type IIB R-R plane-wave background \cite{blau01} using hamiltonian framework was first attempted by Kazama and Yokoi in \cite{kazama08}. In order to formulate the worldsheet theory as a CFT the authors considered Green-Schwarz superstring in semi-light-cone gauge following 
\cite{berkovits04}. The Virasoro algebra was also calculated using a direct method in local form. In this method one directly calculates the commutators $[{\cal T}(\sigma), {\cal T}(\sigma')]$, $[\tilde {\cal  T}(\sigma), \tilde {\cal T}(\sigma')]$ and $[{\cal T}(\sigma), \tilde {\cal T}(\sigma')]$, where ${\cal T}(\sigma)$ and $\tilde {\cal
  T}(\sigma)$ are the right and left moving EM tensor components and
$\sigma$ is the worldsheet space coordinate. The main result of
\cite{kazama08} that is relevant to us may be summarized as
follows\footnote{MNO refers 
  to {\it massless normal ordering} \cite{kazama08}, the one that is relevant to flat
  background. The reason for this nomenclature is that the worldsheet
  theory for flat background may be understood as the massless limit
  of that corresponding to R-R plane-wave in light-cone gauge. PNO was also discussed in the literature earlier (for example \cite{diakonou90}), but we will not go into the technical details of its definition.},     
\bea
&& \hbox{\it If the EM tensor is ordered according to MNO, then the
  Virasoro algebra is not}\cr 
&& \hbox{\it satisfied due to non-zero anomalous terms. However, this anomaly
  vanishes if}\cr
&& \hbox{\it the the EM tensor is ordered with a different prescription called 
    phase-space} \cr 
&& \hbox{\it  normal ordering (PNO).} 
\label{KYresult}
\eea
This result poses a puzzle as defining the EM tensor according to PNO
would imply that the theory does not have a smooth flat-space
limit. This, as we will explain below, is contradictory with the
universality property discussed in \cite{pm0611, pm0704}.

Let us consider an on-shell pp-wave background with flat transverse space
in string theory. The following theorem can be proved (see
\cite{pm0704}) for the corresponding worldsheet theory,
\bea
&&\hbox{\it There exists a universal sector of worldsheet operators
whose correlation functions} \cr
&& \hbox{\it  are evaluated to be same as that in
  flat background.} 
\eea
This implies that the worldsheet theory corresponding to any pp-wave
should admit a smooth flat space limit. In fact it requires that the
spectrum of conformal dimensions within the universal sector of Hilbert space
(i.e. the states created by universal operators) be universal. This
argument implies that PNO can not be the right way to order the EM tensor. 

\subsection{Testing universality argument in bosonic string theory}
\label{ss:bosonic}

In \cite{pm08} we considered the same problem in the simpler
setting of bosonic string theory and showed that the results are
consistent with the above universality argument. It is easy to check
that the PNO ordered EM tensor gives rise to a spectrum of negative
conformal dimensions in flat background. The physical spectrum was
computed for the plane-wave in \cite{pm08} by explicitly constructing the
string field theory (SFT) quadratic term and diagonalizing it. It
indeed shows the existence of negative conformal dimensions for the PNO
ordering. On the other hand the MNO-ordered EM tensor reproduces the
result that is expected from light-cone gauge analysis.  

The Virasoro algebra was also calculated for the off-shell pp-wave
using the same method as in \cite{kazama08}. In addition to the
standard central charge terms, this produces certain additional
operator anomaly terms. The calculations have been done using both PNO
and MNO ordered EM tensor and in both the cases the operator anomaly
terms, ordered accordingly, were shown to be proportional to the
equation of motion for the pp-wave.  

\subsection{A closer look at R-R pp-wave}
\label{ss:RR}

Although the results of \cite{pm08} support the universality argument, the case of R-R pp-wave seems to be more subtle. In particular, the above results do not explain the observation in (\ref{KYresult}). This was explained in \cite{pm0902} which we discuss below.
 
We first write,
\bea
{\cal T} = {\cal T}^{(0)} + \dt {\cal T}~, \quad  
\tilde {\cal T} = \tilde {\cal T}^{(0)} + \dt {\cal T} ~, 
\eea 
where the superscript $(0)$ refers to the flat background, i.e. the
free part and $\dt {\cal T}$ is the interaction part containing
non-trivial background fields. The correct
way of defining the quantum EM tensor was shown to be as follows,
\bea
&& \hbox{\it The free part of the EM tensor is ordered according to
  MNO, while the interaction}\cr 
&& \hbox{\it part needs to be ordered according to
PNO. } 
\label{EM-def}
\eea
The reason for the PNO ordering for the interaction term is as
follows. Given the supergravity background characterized by the
function $K(\vec x) = \int d\vec k~ \tilde K_{sugra}(\vec k) e^{i\vec
  k. \vec x}$, the relevant couplings $\tilde K_{ws}(\vec k)$ in the
worldsheet sigma model are in general related to $\tilde K_{sugra}(\vec k)$ by a field
redefinition. The effect of such field redefinition can be absorbed
into the normal ordering prescription of 
the corresponding interaction term. It turns out that if the term
is ordered according to PNO then the field redefinition becomes
identity. By constructing the SFT quadratic term 
and diagonalizing it one finds that the definition in (\ref{EM-def}) 
reproduces the correct physical spectrum as found in light-cone
analysis \cite{metsaev01}\footnote{The field redefinition discussed above
  involves the transverse laplacian of $K$ which is proportional to the R-R flux
  strength when the background is on-shell. Therefore for the bosonic example considered in
  \cite{pm08}, where there is no R-R flux, the field
  redefinition becomes trivial on-shell. This is why in the bosonic
  case the EM tensor, with both the free and interaction parts ordered according to MNO, gives the right
  spectrum, as mentioned toward the 
  end of subsection \ref{ss:bosonic}.}.   

The explanation of (\ref{KYresult}) goes as follows. The operator
anomaly terms corresponding to the three commutators mentioned above
(\ref{KYresult}) have the following structure,
\bea
{\cal A}^R(\s,\s') &=& A^R(\s,\s') + A_F(\s, \s')~,  \cr
{\cal A}^L(\s,\s') &=& A^L(\s,\s') + A_F(\s, \s')~, \cr
{\cal A}(\s,\s') &=& A(\s,\s') + A_F(\s, \s')~,
\label{anomalies}
\eea
respectively, where $A^R$, $A^L$ and $A$ are the bosonic and $A_F$ is
the fermionic contribution. The calculations are done by introducing a
UV-regulator $\eps$ on the worldsheet. All such terms turn out to have
a special structure: ${\cal O}_{\eps}(\s,\s')\dt_{\eps}(\s-\s')$,
where ${\cal O}_{\eps}(\s,\s')$ is a composite operator anti-symmetric
in $\s$ and $\s'$ and $\dt_{\eps}(\s-\s')$ is the Poisson kernel
representation of Dirac 
delta function which approaches the latter in the limit $\eps \to
0$. Notice that at the classical level, where  
${\cal O}_{\eps}(\s,\s')$ remains the same under reordering of
operators, such a term must 
vanish as expected. In the quantum theory this term is well-defined at
a finite $\eps$. However, first reordering the term in a different way
and then taking $\eps \to 0$ produces a different answer. 

The above result indicates that the direct method of calculating the
Virasoro anomaly suffers from operator ordering ambiguity. It turns
out that if the anomaly terms are ordered according to PNO then the
correct supergravity equation of motion (EOM) is reproduced. This is
the reason for the observation in (\ref{KYresult}). However, there is
no good understanding why such an ordering prescription should work.  

We will come back to this question in section \ref{s:back-indep} where
a background independent construction will be described. In this
construction covariant Ricci-anomaly terms will arise whose technical 
origin is quite different. This indicates that the observation that 
{\it PNO-prescription} described above gives the right
supergravity EOM may be misleading and does not generalize to arbitrary
backgrounds.

\subsection{Dynamical supersymmetry argument for conformal invariance}
\label{ss:susy}

The ordering ambiguity described above may be attributed to the fact that the relevant vacuum is non-perturbative and it is not generically known how to define quantum operators in a controlled manner. However, in this particular case the background is supersymmetric (in fact, as much as flat space) which may be used to gain such control. 
It turns out that in semi-light-cone gauge the following relations hold true in both flat
and R-R plane-wave backgrounds \cite{pm0907},
\bea
 \{Q_{\dot a}, Q_{\dot b} \} = 2 \delta_{\dot a \dot b}  \oint {d\s
   \over 2 \pi} ~\T_{\perp}(\s)~,   \quad
\{\tilde Q_{\dot a}, \tilde
 Q_{\dot b} \} = 2 \delta_{\dot a \dot b}  \oint {d\s \over 2 \pi}
 ~\tilde \T_{\perp}(\s)~,
\label{susy-alg}
\eea
where $Q_{\dot a}$ and $\tilde Q_{\dot a}$ are properly scaled
dynamical supercharges and $\T_{\perp}$ and $\tilde \T_{\perp}$ are
the transverse parts of the EM tensor components. As shown in
\cite{pm0907}, the above relations can be used in certain Jacobi
identities to relate the second order susy variations of the
transverse EM tensor components and certain integrated forms of the
anomaly terms,
\bea
\lt\{[\T_{\perp}(\s), Q_{\dot a}], Q_{\dot b} \rt\} +
\lt\{[\T_{\perp}(\s), Q_{\dot b}], Q_{\dot a} \rt\} &=&
2\delta_{\dot a\dot b}\lt[ \oint {d\s' \over 2\pi} ~{\cal
  A}^R(\s,\s')+ \cdots \rt] ~,\cr 
\lt\{[\tilde \T_{\perp}(\s), \tilde Q_{\dot a}], \tilde Q_{\dot b} \rt\} +
\lt\{[\tilde \T_{\perp}(\s), \tilde Q_{\dot b}], \tilde Q_{\dot a} \rt\} &=&
2\delta_{\dot a\dot b} \lt[ \oint {d\s' \over 2\pi} ~{\cal
  A}^L(\s,\s')+ \cdots \rt] ~, \cr
\lt\{[\tilde \T_{\perp}(\s), Q_{\dot a}], Q_{\dot b} \rt\} +
\lt\{[\tilde \T_{\perp}(\s), Q_{\dot b}], Q_{\dot a} \rt\} &=&
2\delta_{\dot a\dot b} \lt[-\oint {d\s' \over 2\pi} {\cal A}(\s',\s) +
\cdots \rt]~, 
\label{jacobi} 
\eea
where the ellipses refer to certain known terms. The idea is to
compute the anomalies indirectly by calculating the second order susy
variations on the left hand side independently. Although such a
computation does not suffer from ordering ambiguity in the sense
discussed earlier, the problem arise in the form of divergences as  
non-(anti)commuting operators appear at the same point. However, this
subtlety can be taken care of by considering a point-split definition
of the EM tensor. The final results show that all the integrated
anomaly terms appearing on the right hand sides of (\ref{jacobi}) are
zero.  

The above argument establishes conformal invariance only partially as the anomaly terms have been computed with one coordinate integrated over. The complete proof would require one to extend this argument by incorporating local susy currents.

\section{A background independent hamiltonian framework}
\label{s:back-indep}

The susy argument works only for on-shell backgrounds which are
supersymmetric. In order to understand background EOM from Virasoro
anomaly one needs to start with an off-shell background and compute
the Virasoro algebra using the direct method discussed in subsections
\ref{ss:bosonic} and \ref{ss:RR}. However, as discussed earlier this
method leads to ambiguous results for the anomaly terms. In
\cite{pm0912} we explore a background independent framework, hereafter
called DeWitt-Virasoro (DWV) construction, where we attempt to
formulate the problem in a vacuum independent way. Such a framework,
if it exists, should define the worldsheet theory, at least formally, 
over the space of all possible geometrical backgrounds. Starting from 
this framework it should also be possible to derive all the know
properties of the 2-d CFT for a given exact background by specializing 
to it.   

\subsection{DeWitt-Virasoro construction}
\label{ss:DWV}

The approach considered in \cite{pm0912} is to describe the string
quantum mechanics in a coordinate independent way following DeWitt's
argument \cite{dewitt52} in the context of particle quantum
mechanics. 
We consider the simplest case of a bosonic string moving in
an arbitrary metric-background and re-write the worldsheet theory in a
language where it describes a single particle moving in an
infinite-dimensional curved background. The relevant map is given by, 
\bea
x^i =\oint {d\s \over 2\pi}~ X^{\mu}(\s) e^{-im\s} , ~~
g_{ij}(x) = \oint {d\s \over 2\pi} ~G_{\mu \nu}(X(\s))e^{i(m+n)\s}, ~~
a^i(x) =  \oint {d\s \over 2\pi}~ \del X^{\mu}(\s) e^{-im\s},
\label{map}
\eea
where the index identifications are as follows: $i=\{\mu, m\}$,
$j=\{\nu, n\}$ etc. The general coordinate transformation (GCT) in the
physical spacetime: $X^{\mu} \to X'^{\mu}(X)$ induces the same in the
infinite-dimensional spacetime: $x^i \to x'^i(x)$ under which
$g_{ij}(x)$ transforms as metric tensor and $a^i(x)$ as a vector
field. The classical worldsheet lagrangian takes the following form in
terms of the new set of variables,
\bea
L(x, \dot x) &=& {1\over 2} g_{ij}(x) \lt[\dot x^i \dot x^j -  a^i(x)
a^j(x) \rt]~.
\label{L}
\eea

GCT is a point canonical transformation which can be described by a
unitary transformation in the quantum theory. DeWitt's analysis in
\cite{dewitt52} shows how to construct the quantum mechanics such that
general covariance is manifest in its position space representation. As a
result it gives generally covariant expressions for all the matrix
elements.  

Following DeWitt's method we formally define a background
independent notion of the quantum Virasoro generators and algebra,
hereafter called DWV generators and algebra. We first write down the
classical Virasoro generators using the particle language which then
leads to the following natural definition (consistent with the
hermiticity property) of the quantum DWV 
generators, 
\bea
L_{(i)} &=& {1\over 4} \lt[ \pih^{\star}_k g^{k l+i}(\xh) \pih_l -
(\pih^{\star}_k a^{k+i}(\xh) + a^{k+i}(\xh) \pih_k ) +g_{kl}(\xh)
a^k(\xh) a^{l+i}(\xh) \rt]~, \cr 
\tilde L_{(i)} &=& {1\over 4} \lt[ \pih^{\star}_k g^{k l+\bar i}(\xh)
\pih_l +  (\pih^{\star}_k a^{k+\bar i}(\xh) + a^{k+\bar i}(\xh) \pih_k
) +g_{kl}(\xh) a^k(\xh) a^{l+\bar i}(\xh) \rt]~, 
\label{DWV-gen}
\eea
where given the index identification below eqs.(\ref{map}),  we have
defined: $\bar i = \{\mu, -m\}$, $(i)=m$ and $i+j=\{\mu, m+n\}$. The
$\hat \pi$-operators are given by (in $\hbar=\alpha'=1$ unit) \cite{omote72}, 
\bea
\pih_j = \ph_j + {i\over 2} \gamma_j(\xh)~, \quad \pih_j^{\star} =
\ph_j - {i\over 2} \gamma_j(\xh)~, 
\label{pih-def}
\eea
where $\gamma_j=\gamma^k_{jk}$ are the contracted Christoffel
symbols. DeWitt's technique in \cite{dewitt52} allows us to calculate
the matrix elements in spin-zero representation. We find that the DWV
algebra in spin-zero representation is given by the Witt algebra with
additional anomaly terms that vanish for Ricci-flat backgrounds, 
\bea
\la \chi| \lt\{ \begin{array}{l}
[\hat L_{(i)}, \hat L_{(j)}] = (i-j) \hat L_{(i+j)} \cr
[\hat{\tilde L}_{(i)}, \hat{\tilde L}_{(j)}] = (i-j) \hat{\tilde L}_{(i+j)} \cr
[\hat L_{(i)}, \hat{\tilde L}_{(j)}] = \displaystyle{{1\over 8}
  \lt(\hat \pi^{\star k+i}  
r_{kl}(\hat x) a^{l+\bar j}(\hat x) - a^{k+i}(\hat x) r_{kl}(\hat x)
\hat \pi^{l+\bar j} \rt)} 
\end{array} \rt\} |\psi \ra ~,
\label{scalar-alg}
\eea
where $|\chi \ra$ and $|\psi \ra$ are two arbitrary spin-zero states
and $r_{ij}$ is the Ricci tensor. 

The DWV generators constructed above are background independent and
are not normal ordered with respect to any particular vacuum. However,
as shown in \cite{pm0912}, the quantum Virasoro generators in flat and
pp-wave backgrounds can be obtained by normal ordering the DWV
generators in a suitable way. The resulting quantum Virasoro algebra
contains the central charge terms in addition to the operator anomaly
terms appearing in (\ref{scalar-alg}). This enables us to compare the
(ambiguous) operator anomaly terms in the bosonic sector  
in (\ref{anomalies}) with the one in (\ref{scalar-alg}), at least in
the spin-zero representation. This comparison tells us that all the
operator anomaly terms $A^R$, $A^L$ and $A$ in (\ref{anomalies}) must
vanish. Furthermore it also implies that the Ricci-term in
(\ref{scalar-alg}) was missing in the previous computation. This
apparent discrepancy can be resolved by the following observation. By
going to the position space representation 
it can be argued that the matrix element of the Ricci-term in
(\ref{scalar-alg}) vanishes, though the Ricci tensor itself does not,
because of certain restrictions on the index contraction for pp-wave
and that the Ricci tensor depends only on the transverse coordinates.  

\subsection{Higher rank tensor generalization}
\label{ss:tensor}

An important question for the construction described above is how to
generalize this to higher spin representations which are very relevant
in string theory. In particular, one would like to know if the DWV
algebra in (\ref{scalar-alg}) holds true in arbitrary spin
representations. We studied this question in \cite{pm10} and found the
answer in affirmative.  

In flat background one usually obtains the higher rank tensor states
by applying suitable creation operators on the ground state. It turns
out that such state-operator mapping is not suitable for the
background independent framework. What turn out to be more relevant in
this case are the Sch$\ddot{\rm{o}}$dinger wavefunctions. Such
wavefunctions can be obtained either by multiplying a tensor field
with a scalar wavefunction or by applying covariant derivatives on it.   

In \cite{pm10} we consider a new framework where an arbitrary tensor
state of rank $n$, which is by itself coordinate invariant, is
expanded as follows, 
 \bea
|\psi_{(n)}\ra = \int dw~ \psi^{i_1 i_2\cdots i_n}(x)
|{}_{i_1i_2\cdots i_n} ; x\ra ~, 
\label{psi-n}
\eea
where $\psi^{i_1 i_2\cdots i_n}(x)$ is the tensor wavefunction under
consideration and $|{}_{i_1i_2\cdots i_n} ; x\ra$ 
are the rank-$n$ position eigenstates which are newly introduced. 
The problem then reduces to finding the representation of
the momentum operator in such basis states. This is done by suitably
generalizing DeWitt's argument in \cite{dewitt52}. The result is given
by, 
\bea
\la{}_{i_1\cdots i_n};x|\pih_k|{}_{j_1\cdots j_{\tilde n}};\tilde x\ra
&=& -i \dt_{n,\tilde n} \lt[\Delta_{i_1\cdots i_n j_1\cdots j_n}(x,\tilde x)
\del_k\dt(x,\tilde x)  + F_{i_1\cdots i_n j_1\cdots j_n k}(x,\tilde x)
\dt(x,\tilde x)\rt] ~, \cr && 
\label{pi-tensor-rep}
\eea
where $\del_k \equiv {\del \over \del x^k}$,
\bea
\dt(x,\tilde x) = {\dt(x-\tilde x) \over  \lt(g(x) g(\tilde
  x)\rt)^{{1\over 4}}}~, 
\label{delta}
\eea
$\dt(x-\tilde x)$ being the Dirac delta function, $g =|{\rm det}g_{ij}|$,
\bea
F_{i_1\cdots i_n j_1\cdots j_n k}(x,\tilde x) &=& -\nabla_k \Delta_{i_1\cdots  i_n j_1\cdots j_n}(x,\tilde x) ~,
\label{F-n-def}
\eea
where $\nabla_k$ is the covariant derivative with respect to $x^k$ and
\bea
\Delta_{i_1\cdots i_n j_1\cdots j_n}(x,\tilde x) = \Delta_{i_1j_1}(x,\tilde x)
\cdots \Delta_{i_nj_n}(x,\tilde x) ~.
\label{Delta-n}
\eea
To define $\Delta_{ij}(x,\tilde x)$ we proceed as follows. Let us
first consider two points $X^{\mu}$ and $\tilde X^{\mu}$ in the
physical spacetime sufficiently close to each other so that a unique geodesic passes
through them. Then following \cite{dewitt60} we define the bi-vector
of geodetic parallel transport $D_{\mu \nu}(X, \tilde X)$.  
The two vector-indices $\mu$ and $\nu$ are associated to the two
points $X$ and $\tilde X$ respectively such that under a GCT the
bi-vector transform as a vector at each point separately. Moreover,
$D_{\mu \nu}(X, \tilde X)$, contracted with another vector at one of
the points, gives the same vector parallel transported to the other
point along the geodesic. One generalizes this to a string by
considering two string embedding $X^{\mu}(\s)$ and $\tilde
X^{\mu}(\s)$ such that two points belonging to the two strings at the
same value of $\s$ are connected by a unique geodesic. This gives us
the parallel transport bi-vector $D_{\mu \nu}(X(\s), \tilde X(\s))$ on
the worldsheet whose infinite-dimensional counterpart is given
by $\Delta_{ij}(x,\tilde x)$, 
\bea
\Delta_{ij}(x,\tilde x) &=& \oint {d\s \over 2\pi}~ D_{\mu \nu}(X(\s), \tilde X(\s))
e^{i(m+n)\s} ~.
\label{Deltaij}
\eea
The above definition can be generalized to any other well-behaved
paths which are not necessarily geodesics. However, the expression in
(\ref{pi-tensor-rep}), being sensitive to the expansion of
$\Delta_{ij}(x,\tilde x)$ only up to first order in separation, is
insensitive to the choice of path. 

Although the expression in (\ref{pi-tensor-rep}) makes the tensorial
properties of the relevant matrix element manifest\footnote{Under a GCT, $\hat x^i\to \hat x'^i(\hat x)$: $\hat \pi_k \to \hat \pi'_k = \lambda_k^{~l}(\hat x) \hat \pi_l$, where $\lambda_k^{~l}(\hat x)$ is the inverse of the Jacobian matrix for the transformation. This implies that the matrix element on the left hand side of (\ref{pi-tensor-rep}) has a covariant vector index $k$ at $x$. This is obviously true for the second term on the right hand side. It is also true for the first term as  $\dt(x,\tilde x)$ is a bi-scalar.}, the actual
calculations are done using an explicit representation of
$\Delta_{ij}(x,\tilde x)$ given in terms of the vielbeins. Given this
basic ingredient, one can compute the DWV generators in tensor
representations. These expressions contain all the terms that are
present in the scalar representation. Moreover, they receive
additional contributions involving spin connection. 

Finally, following \cite{pm0912} we proceed to compute the DWV
algebra. As expected, the computation gets more complicated than the
previous one because of the additional contributions. However, we show
that all the unwanted contributions cancel in various ways to
reproduce the same algebra in (\ref{scalar-alg}). There arise three
kinds of additional terms which depend on Riemann tensor, covariant
derivative of spin connection and higher power of spin connection. It
turns out that because of the particular definition in
(\ref{DWV-gen}) these three kinds of terms are organized in such a way
that they cancel due to various identities in general relativity.

\section{Final remarks}
\label{s:final}

The fact that the same Ricci-flatness condition is obtained as the
condition of vanishing anomaly in the DWV algebra in arbitrary tensor
representations is interesting. An important question that arises
at this point is how to understand the conventional CFT's as
special cases of the DWV construction. This is understood for flat and
pp-wave backgrounds in spin-zero representation. However, because of
the new way of describing the tensor states as in (\ref{psi-n}) it is
interesting to explore how exactly the conventional CFT computations
should be realized in our background independent description. It would
also be interesting to understand the central charge terms in the
Virasoro algebra in a background independent way. We hope to come back
to these questions in future.

\ack{I thank Sumit R. Das and Alfred Shapere, the organizers of the symposium 
Quantum Theory and Symmetries 6, for such an exciting event.}

\end{document}